\documentclass[letterpaper,11pt]{article}
\usepackage[margin=1in]{geometry}
\usepackage[utf8]{inputenc} 
\usepackage[T1]{fontenc}    
\usepackage[colorlinks=true,breaklinks=true,bookmarks=true,urlcolor=blue,citecolor=blue,linkcolor=blue,bookmarksopen=false,draft=false]{hyperref}
\usepackage{url}            
\usepackage{booktabs}       
\usepackage{multirow}
\usepackage{microtype}      
\usepackage{amsfonts,amsmath, amsthm,mathtools,stmaryrd}
\usepackage{xcolor}
\usepackage[inline]{enumitem}
\usepackage{appendix}
\usepackage{float, graphicx, subcaption}

\newcommand{\X}{\mathcal X}

\def\argmin_#1{\underset{#1}{\mathrm{arg\,min\, }}}
\def\argmax_#1{\underset{#1}{\mathrm{arg\,max\, }}}

\makeatletter
\def\dasharrowfill@#1#2#3#4{%
        $\m@th
        \thickmuskip0mu
        \medmuskip\thickmuskip
        \thinmuskip\thickmuskip
        \relax
        #4#1\mkern2mu
        \xleaders\hbox{$#4\mkern2mu#2\mkern2mu$}\hfill
        \mkern2mu
        #3$%
}

\def\dashleftarrowfill@{\dasharrowfill@\leftarrow\relbar\relbar}
\def\dashrightarrowfill@{\dasharrowfill@\relbar\relbar\rightarrow}
\def\dashleftrightarrowfill@{\dasharrowfill@\leftarrow\relbar\rightarrow}
\def\dashLeftarrowfill@{\dasharrowfill@\Leftarrow\Relbar\Relbar}
\def\dashRightarrowfill@{\dasharrowfill@\Relbar\Relbar\Rightarrow}
\def\dashLeftrightarrowfill@{\dasharrowfill@\Leftarrow\Relbar\Rightarrow}

\providecommand*\xdashleftarrow[2][]{%
  \ext@arrow 0055{\dashleftarrowfill@}{#1}{#2}}
\providecommand*\xdashrightarrow[2][]{%
  \ext@arrow 0055{\dashrightarrowfill@}{#1}{#2}}
\providecommand*\xdashleftrightarrow[2][]{%
  \ext@arrow 0055{\dashleftrightarrowfill@}{#1}{#2}}
\providecommand*\xdashLeftarrow[2][]{%
  \ext@arrow 0055{\dashLeftarrowfill@}{#1}{#2}}
\providecommand*\xdashRightarrow[2][]{%
  \ext@arrow 0055{\dashRightarrowfill@}{#1}{#2}}
\providecommand*\xdashLeftrightarrow[2][]{%
  \ext@arrow 0055{\dashLeftrightarrowfill@}{#1}{#2}}
\makeatother

\begin{document}

\title{Transfer Learning for Portfolio Optimization}

\author{
Haoyang Cao
\thanks{Centre de Math\'ematiques Appliqu\'ees, Ecole Polytechnique.
\textbf{Email:}
haoyang.cao@polytechnique.edu}
\and
Haotian Gu
\thanks{Department of Mathematics, UC Berkeley.
\textbf{Email:} 
haotian{\textunderscore}gu@berkeley.edu }
\and
Xin Guo
\thanks{Department of Industrial Engineering \& Operations Research, UC Berkeley.
\textbf{Email:} 
xinguo@berkeley.edu}
\and
Mathieu Rosenbaum
\thanks{Centre de Math\'ematiques Appliqu\'ees, Ecole Polytechnique.
\textbf{Email:}
mathieu.rosenbaum@polytechnique.edu}
}
\date{July 25, 2023}
\maketitle

\begin{abstract}
    In this work, we explore the possibility of utilizing transfer learning techniques to address the financial portfolio optimization problem. We introduce a novel concept called ``transfer risk'',  within the optimization framework of transfer learning. A series of numerical experiments are conducted from three categories: cross-continent transfer, cross-sector transfer, and cross-frequency transfer. In particular, 
    \begin{enumerate*}
        \item a strong correlation between the transfer risk and the overall performance of transfer learning methods is established, underscoring the significance of transfer risk as a viable indicator of ``transferability'';
        \item transfer risk is shown to provide a computationally efficient way to identify appropriate source tasks in transfer learning, enhancing the efficiency and effectiveness of the transfer learning approach;
        \item additionally, the numerical experiments offer valuable new insights for portfolio management across these different settings.  
    \end{enumerate*}
\end{abstract}

\section{Introduction}
\subsection{Transfer Learning}
Transfer learning is built upon the fundamental principle that knowledge gained from solving one problem can be transferred and applied to effectively solve a different but related problem. As a powerful technique in machine learning, it has attracted considerable attention and showcased remarkable potential across various domains, including natural language processing \cite{ruder2019transfer, devlin-etal-2019-bert, sung2022vl}, sentiment analysis \cite{jiang2007instance, deng2013sparse, liu2019survey}, computer vision \cite{deng2009imagenet, long2015learning, ganin2016domain, wang2018deep}, activity recognition \cite{cook2013transfer, wang2018stratified}, medical data analysis \cite{zeng2019automatic, wang2022transfer, kim2022transfer}, and bio-informatics \cite{hwang2010heterogeneous}. It is also one of the backbones for large language models such as the GPT-series \cite{ouyang2022training,openai2023gpt4}. See also review papers \cite{survey1,tan2018survey,zhuang2020comprehensive}. 

Apart from the above practical adoption of transfer learning, there has been a line of work dedicating to its theoretical aspects. Some of them tend to focus on specific learning problems, such as classification, and derive upper bounds of generalization error under different measurements. There are the VC-dimension of the hypothesis space adopted in \cite{blitzer2007learning},  total variation distance in \cite{ben2010theory}, $f$-divergence in \cite{harremoes2011pairs}, Jensen-Shannon divergence in \cite{zhao2019learning}, $\mathcal{H}$-score in \cite{bao2019information}, mutual information in \cite{bu2020tightening}, and more recently $\X^2$-divergence in \cite{tong2021mathematical}, and variations of optimal transport cost in \cite{tan2021otce}. Others interpret transferability for transfer learning as a measurement of similarity between the source and the target data using various divergences, such as low-rank common information in \cite{saenko2010adapting}, KL-divergence in \cite{ganin2015unsupervised,ganin2016domain,tzeng2017adversarial}, $l_2$-distance in \cite{long2014transfer}, and the optimal transport cost in \cite{flamary2016optimal}. 

\subsection{Transfer Learning in Finance}
In the realm of finance, limited data availability or excessive noise can hinder practitioners from accomplishing tasks such as equity fund recommendation \cite{zhang2018equity} and stock price prediction \cite{wu2022jointly}. Under this circumstance, transfer learning offers a promising avenue for overcoming data constraints and improving predictive models. Instead of starting from scratch for each specific task, transfer learning allows financial practitioners to capitalize on the knowledge and patterns accumulated from analogous tasks or domains. By transferring the knowledge, models can effectively learn from past experiences and generalize to new situations, resulting in more accurate predictions and enhanced decision-making capabilities. For instance, in \cite{zhang2018equity}, the investment strategy gained from the stock market is transferred to the equity market where data are unconventional in order to facilitate personalized investment recommendation. In \cite{leal2020learning}, to overcome the scarcity of training data in high frequency trading, the deep learning model designed for trading strategy generation is first pretrained over simulated data before fine-tuning over genuine historical trading trajectories. In \cite{wu2022jointly}, to improve the accuracy of stock prediction, the knowledge of industrial chain information is transferred to the prediction model. In fact, to assist in stock and market prediction, there has been a stream of work utilizing the advances in natural language processing to extract useful information from financial text. One such example would be FinBERT \cite{liu2021finbert}, a financial text mining variant of the BERT model. For more examples and details, see survey papers on natural language based financial forecasting such as \cite{xing2018natural}. Transfer learning techniques can also be applied to other applications of finance and economics including credit risk management \cite{lebichot2020deep}, model calibration \cite{rosenbaum2021deep}, and crude oil price prediction \cite{cen2019crude}. 

\subsection{Our Work}
In this work, we will apply {\em transfer learning} techniques to the {\em  portfolio optimization} problem. Moreover, we will introduce a new concept called {\em transfer risk}  within the optimization framework for transfer learning established in \cite{CGG2023}, and demonstrate its connection with the learning outcome  via different numerical experiments.  Numerical evidence shows a strong correlation between the transfer risk and the overall performance of transfer learning methods, indicating the significance of transfer risk as a viable indicator of the transferability. Moreover, it shows that transfer risk provides a computationally efficient way to identify appropriate source tasks in transfer learning, enhancing the efficiency and effectiveness of the transfer learning approach.

For the experiments, we test the performance of transfer learning and transfer risk under three tasks, namely, cross-continent transfer, which is transferring a portfolio from the US equity market to the other equity markets, cross-sector transfer, which is transferring a portfolio from the one sector to other sectors, and cross-frequency transfer, which is transferring a low-frequency portfolio to the mid-frequency domain. 
The numerical experiments offer valuable insights into the potential of transfer learning across these different settings. 
\begin{itemize}
    \item 
In the cross-continent transfer, our study shows different performance for different international market. For instance,  transfer learning from the US market outperforms direct learning in certain European markets such as Germany, but it performs relatively poor for the Brazil market. This suggest that Germany market is a better candidate as a transfer target than the Brazil market.
\item For the cross-sector transfer, we analyze the performance differences among sectors, and discuss the underlying factors. For instance, our analysis reveals that transfer risks in Health Care and Information Technology sectors display large negative correlations. In contrast, correlations are not significant for Utilities and Real Estate, potentially due to factors not fully captured by transfer risk.
\item Regarding the cross-frequency transfer, our results indicate that transferring a low-frequency portfolio (one-day) to higher frequencies (intraday) carries  high transfer risks with poor performances. In contrast, transferring within the mid to high-frequency regime yields more robust and promising outcomes.
\end{itemize}
\subsection{Paper outline} In Section \ref{sec:methods}, we will formally define the transfer learning framework of the financial portfolio optimization problem. Based on this optimization framework, we will define a new concept called {\em transfer risk} to provide an {\it apriori} estimate on the final learning outcome. In Section \ref{sec:experiments}, we will state the settings for the numerical tests and present corresponding results and implications. Finally, we conclude this work by Section \ref{sec:conclusion}.

\section{Portfolio Optimization Problem and Transfer Learning}\label{sec:methods}

\subsection{Problems}
In many cases such as managing portfolios in new emerging markets, there are limited data for directly estimating the mean return and covariance matrix for a set of portfolio, resulting in large estimation error and non-robust portfolio. We will
show that transferring knowledge from portfolios of a mature market could be a natural and viable way to tackle this problem. 
The basic idea of transfer learning is simple: it is  to  leverage knowledge from a well-studied portfolio optimization problem in a mature market, known as the source task, to improve the performance of a new portfolio optimization  problem in the emerging market with similar features, known as the target task. 
In particular, we will consider three types of portfolio transfers: 

\begin{enumerate}
    \item Cross-continent transfer: This refers to the case when one transfers a portfolio from the equity market in one country to the equity market in another, e.g., from the US equity market to the Brazil equity market. In general, the source market has more historical data or more diverse stocks than the target market, which may provide the target market with a robust pre-trained portfolio. The study of cross-continent transfer aims to understand how continental discrepancy affects the performance of transfer learning. 

    \item Cross-sector transfer: This refers to the case when one transfers a portfolio from one sector of a market to another sector, e.g., from Information Technology sector to Health Care sector in the US equity market. The study of cross-sector transfer aims to understand correlations between various sectors in the market, and how correlations between sectors affect the performance of transfer learning.

    \item Cross-frequency transfer: This refers to the case when one transfers a portfolio constructed under one trading frequency to another trading frequency, e.g., from low-frequency trading to mid or high-frequency trading. The study of cross-frequency transfer aims to explore the possibility of transferring the portfolio across different trading frequencies, which has important and practical implications for institutional investors.
\end{enumerate}

\subsection{Mathematical Setup}\label{subsec:port_opt_setup}
First, let us start by considering a capital market consisting of $d$ assets whose annualized returns are captured by the random vector $r = (r_1, ..., r_d)^\top \sim \mathbb{P}.$ A portfolio allocation vector $\phi = (\phi_1, ..., \phi_d)^\top$ is a $d$-dimensional vector in the unit simplex $\mathbb{X} := \{\phi \in \mathbb{R}^d_+: \sum_{j=1}^d \phi_j = 1 \}$ with $\phi_i$ percentage of the available capital invested in asset $i$ for each $i=1,...,d$. The annualized return of a portfolio $\phi$ is given by $\phi^\top r.$ 

Recall that the optimal portfolio problem  is to find  the highest Sharpe ratio from solving the following optimization problem:    
\begin{equation}\label{eqn:portfolio_opt}
    \widehat{\phi}=\argmax_{\phi \in \mathbb{X}}\frac{\mathbb{E}^{\mathbb{P}}[\phi^\top r]}{\text{Var}(\phi^\top r)}=\argmax_{\phi \in \mathbb{X}}\frac{\mu_\mathbb{P}^\top\phi}{\phi^\top\Sigma_\mathbb{P}\phi},
\end{equation}
where $\mu_\mathbb{P}$ is the expectation and $\Sigma_\mathbb{P}$ is the covariance matrix  of the return $r$. Empirically, $\mu_\mathbb{P}$ and $\Sigma_\mathbb{P}$ are estimated from the historical return.

 Next, we will denote the source task as \(S\) and target task as \(T\) for the problem of  portfolio optimization, and assume that the source and target tasks \(S\) and \(T\) share the same input and output spaces denoted by $\mathcal{X}_S=\mathcal{X}_T=\mathbb{R}^d$ and $\mathcal{Y}_S=\mathcal{Y}_T=\mathbb{R}$. Note that different from other financial machine learning problems, such as return prediction, there is no sample from the output space that is explicitly observed. Instead, one can only get historical stock returns as input data. Meanwhile, a portfolio vector $\phi$ can be viewed as a linear mapping from $\mathbb{R}^d$ to $\mathbb{R}$, taking a $d$-dimensional stock return to a portfolio return. More specifically, the admissible sets of source and target models are restricted to: $A_S=A_T=\{f:\mathbb{R}^d\to\mathbb{R}|f(r)=\phi^\top r\text{ for some }\phi\in\mathbb{X}\}$. In addition, for the source task, the loss functional is set to be the negative Sharpe ratio, and consequently, the source task is to solve following the optimization problem:
\begin{equation}\label{eqn:portfolio_opt_source}
    \widehat{\phi_S}=\argmax_{\phi \in \mathbb{X}}\frac{\mu_S^\top\phi}{\sqrt{\phi^\top\Sigma_S\phi}},
\end{equation}
where $\mu_S$ and $\Sigma_S$ are the mean and covariance estimations from the source data set. 

\begin{figure}
    \centering
    \includegraphics[width=0.7\textwidth]{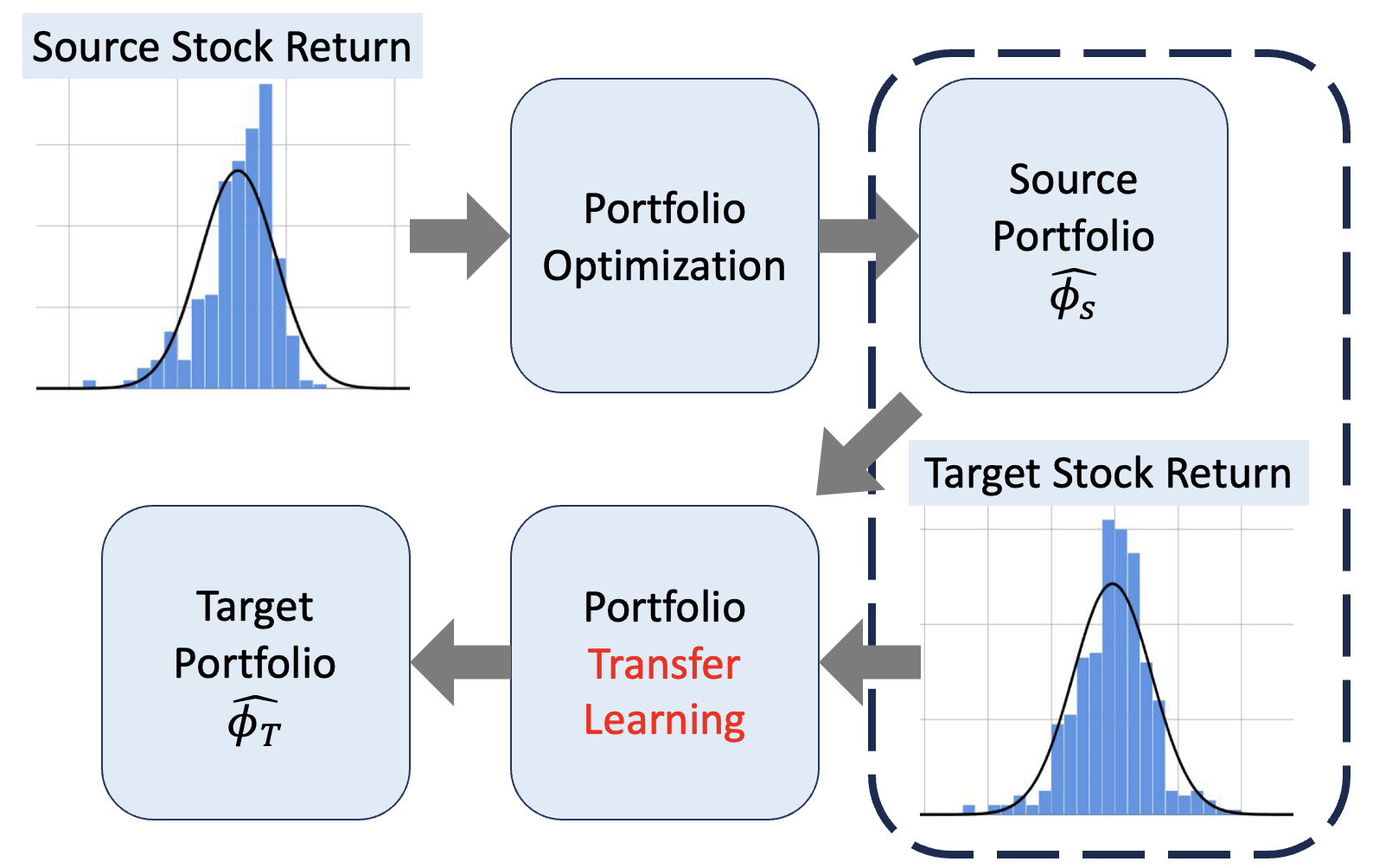}
    \caption{Procedure of portfolio transfer.}
    \label{fig:procedure}
\end{figure}

By the transfer learning approach, we will transfer the portfolio optimization problem from the source task to the target task via solving the following optimization with a $L_2$ regularization term penalizing the distance between the pre-trained portfolio and the transferred portfolio:
\begin{equation}\label{eqn:portfolio_opt_target}
    \widehat{\phi_T}=\argmax_{\phi \in \mathbb{X}}\frac{\mu_T^\top\phi}{\sqrt{\phi^\top\Sigma_T\phi}}-\lambda\left\|\widehat{\phi_S}-\phi\right\|^2_2.
\end{equation}
Here $\mu_T$ and $\Sigma_T$ are the mean and covariance estimations from the target data, and $\lambda>0$ is a hyper-parameter controls the power of the regularization: the higher $\lambda$ is, the closer the transferred portfolio $\widehat{\phi_T}$ will be to the pre-trained portfolio $\widehat{\phi_S}$. In fact, \eqref{eqn:portfolio_opt_target} is equivalent to searching an output transport mapping $T_Y$ over the linear function space $A_T=\{f:\mathbb{R}^d\to\mathbb{R}|f(r)=\phi^\top r\text{ for some }\phi\in\mathbb{X}\}$, while the loss functional takes the form of negative Sharpe ratio in addition to a regularization on the $l_2$ distance from the source model $\widehat{\phi_S}$. The whole procedure of the portfolio transfer is summarized in Figure \ref{fig:procedure}.

To assess the risk of transfer learning in the portfolio optimization problem, we  will devise a risk metric called {\em transfer risk} denoted by $R_\text{trans}$.
It encompasses two aspects we take into consideration, namely the ``quality'' and the ``relevance'' of the chosen source portfolio. 
That is,  the transfer risk $R_\text{trans}$ is expressed as
\begin{equation}
R_\text{trans}=R_1+R_2.
\end{equation}
Here $R_1$  concerns with the performance of the source portfolio, and is defined as
\begin{equation}
R_1 = \left(\frac{\mu_S^\top\widehat{\phi_S}}{\sqrt{\widehat{\phi_S}^\top\Sigma_S\widehat{\phi_S}}}\right)^{-1}.
\end{equation}
This expression is inversely proportional to the Sharpe ratio of the source task, and it serves as a measure of risk associated with selecting a source task that exhibits poor portfolio performance.

The second component $R_2$ measures the similarity between the source and target portfolios in terms of their data distributions. More specifically, it approximate the return distributions of the source and target tasks using mean and covariance estimates $\left(\mu_S, \Sigma_S\right)$ and $\left(\mu_T, \Sigma_T\right)$, respectively, and  model these distributions as multivariate normal distributions $\mathcal{N}\left(\mu_S, \Sigma_S\right)$ and $\mathcal{N}\left(\mu_T, \Sigma_T\right)$. Consequently, $R_2$ is defined as the Wasserstein-2 distance between these two distributions:
\begin{equation}
R_2=\mathcal{W}_2\left(\mathcal{N}\left(\mu_S, \Sigma_S\right), \mathcal{N}\left(\mu_T, \Sigma_T\right)\right).
\end{equation}

\section{Experimental Settings and Results}\label{sec:experiments}
Throughout the experiments, we test the performance of transfer learning and transfer risk under various portfolio optimization problems, including cross-continent transfer, cross-sector transfer, and cross-frequency transfer.
\subsection{Cross-Continent Transfer.} 
In these numerical experiments, the source market is defined as the US equity market, while the target markets are chosen to be United Kingdom, Brazil, Germany, and Singapore, in four separate experiments respectively. Our findings indicate that transfer learning is more likely to outperform direct learning in European markets, such as Germany, while its performance is comparatively worse in the Brazil market. Notably, we observe a strong correlation between transfer risk and the transfer learning performance across all the different markets.

Given a target market, we first select out the top ten stocks with the largest market capitals ($d=10$) as the class of target assets. Then, ten stocks will be randomly selected from the S\&P500 component stocks as the class of source assets. Three data sets will be constructed accordingly:
\begin{enumerate}
    \item \textbf{Source training data}: it consists of the daily returns of ten source assets, from February 2000 to February 2020.

    \item \textbf{Target training data}: it consists of the daily returns of ten target assets, from February 2015 to February 2020.

    \item \textbf{Target testing data}: it consists of the daily returns of ten target assets, from February 2020 to September 2021.
\end{enumerate}

We compare direct learning with transfer learning: for direct learning, the portfolio is directly learned by solving \eqref{eqn:portfolio_opt}, with mean and covariance estimated from target training data; for transfer learning, the portfolio is first pre-trained on the source training data by solving \eqref{eqn:portfolio_opt_source}, then fine-tuned on the target training data by solving \eqref{eqn:portfolio_opt_target}. Finally, the performances of those methods are evaluated through their Sharpe ratios on the target testing data. The regularization parameter $\lambda$ in \eqref{eqn:portfolio_opt_target} is set to be $0.2$. Meanwhile, we also compute the transfer risk following the method described in Section \ref{subsec:port_opt_setup}, using the source training data and target testing data.

\begin{figure}[!ht]
     \centering
     \begin{subfigure}[b]{0.4\textwidth}
         \centering
         \includegraphics[width=\textwidth]{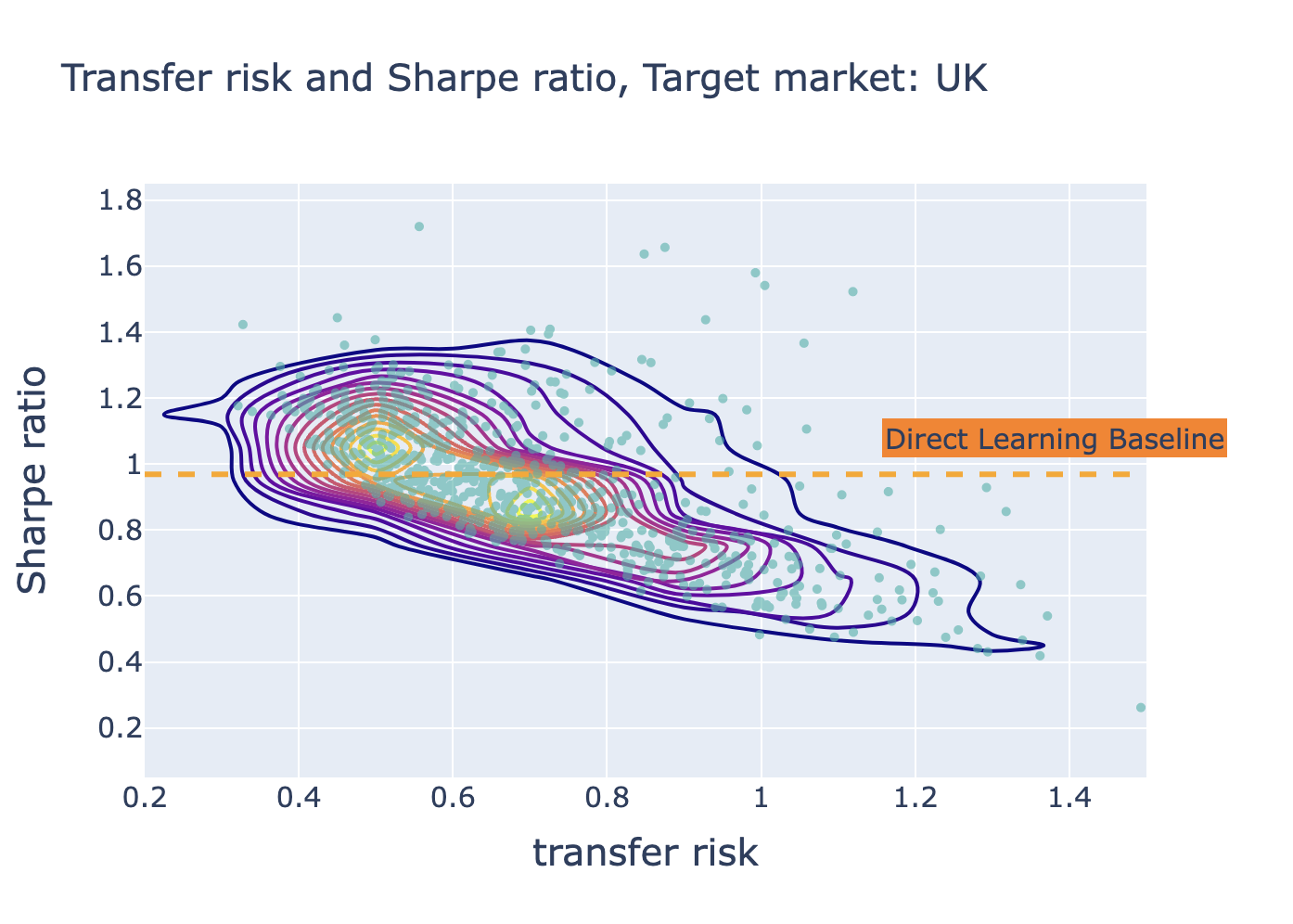}
         \caption{US$\to$UK}
         \label{fig:US-UK}
     \end{subfigure}
     ~
     \begin{subfigure}[b]{0.4\textwidth}
         \centering
         \includegraphics[width=\textwidth]{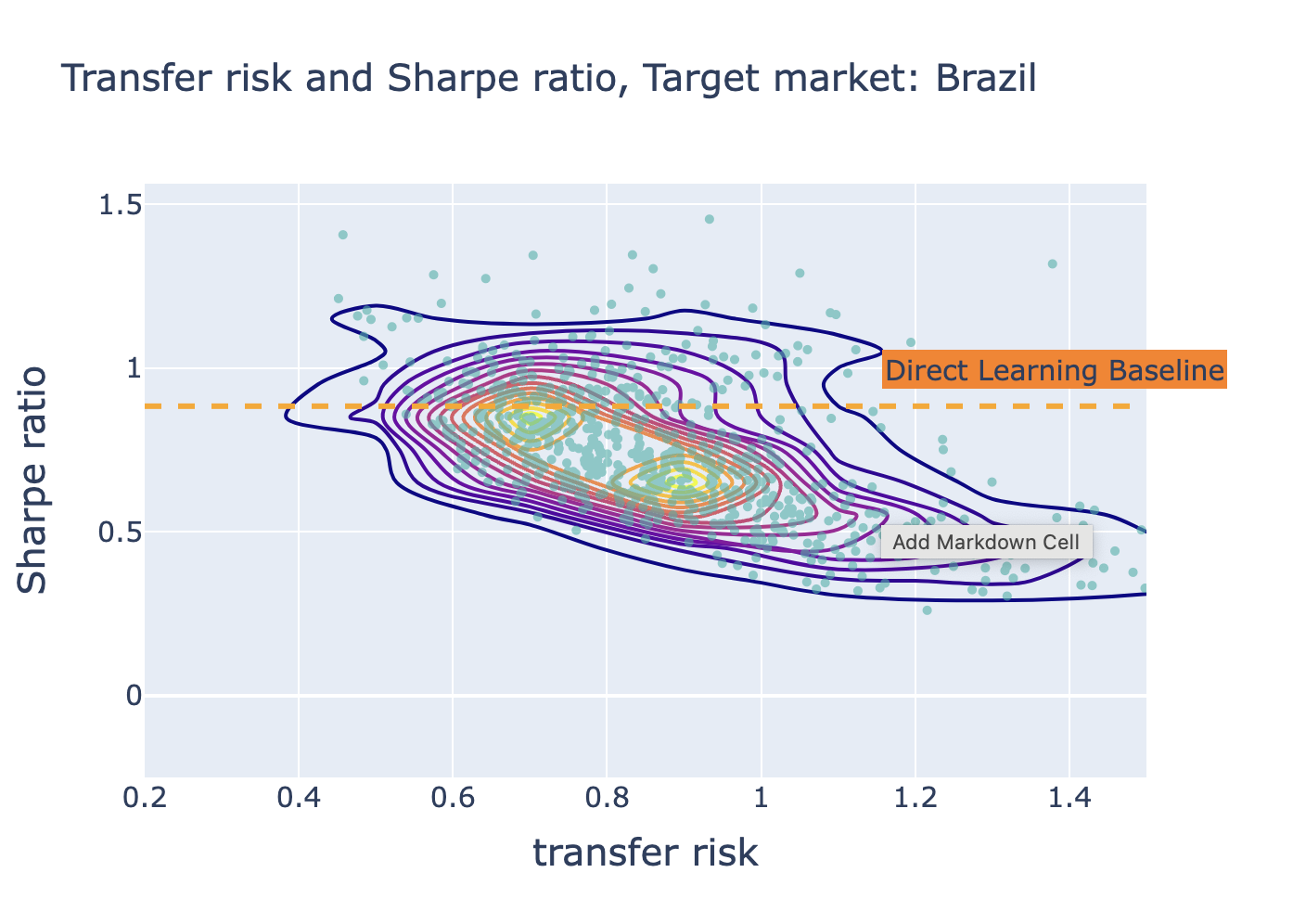}
         \caption{US$\to$Brazil}
         \label{fig:US-Brazil}
     \end{subfigure}\\
     \begin{subfigure}[b]{0.4\textwidth}
         \centering
         \includegraphics[width=\textwidth]{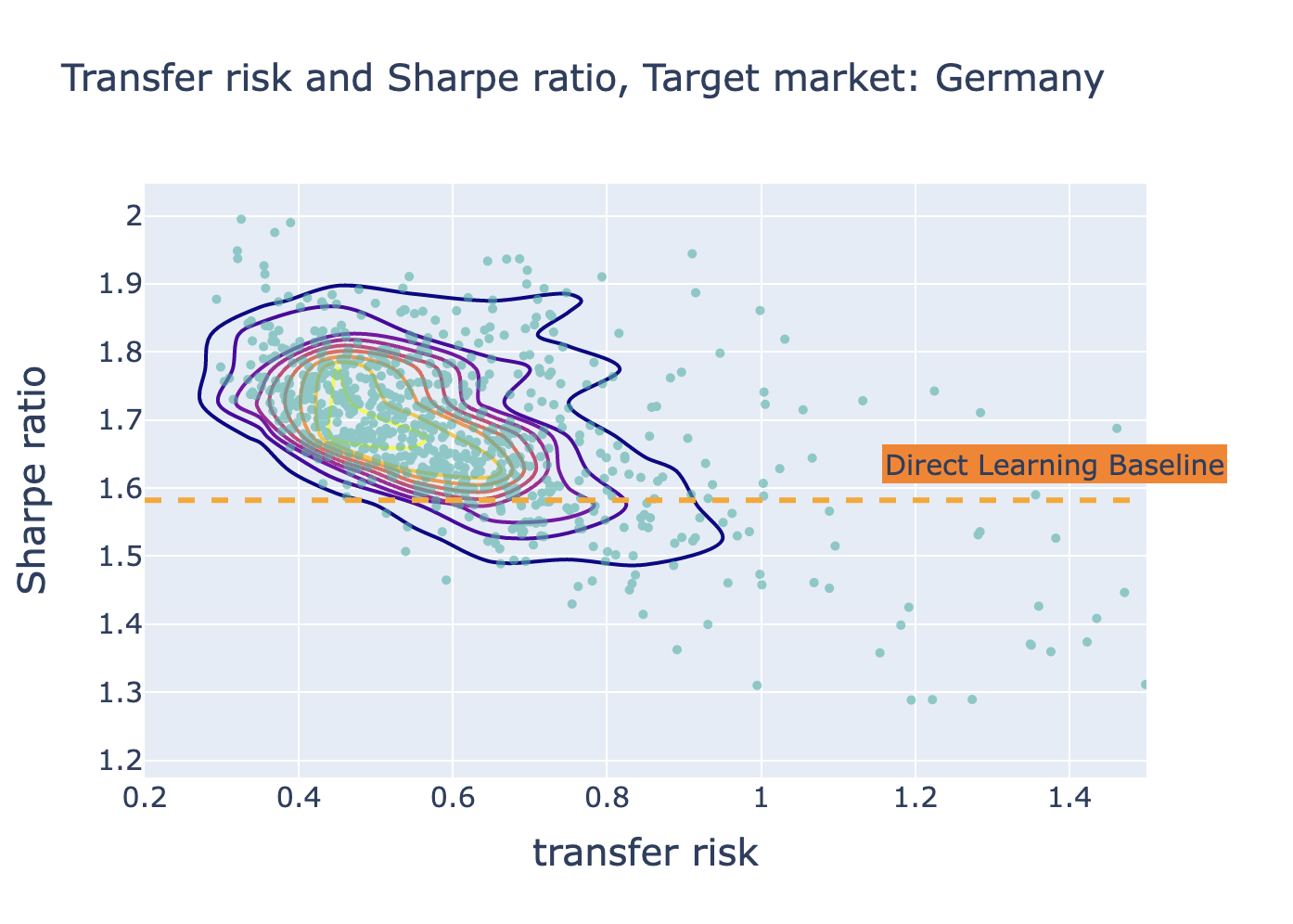}
         \caption{US$\to$Germany}
         \label{fig:US-Germany}
     \end{subfigure}
     ~
     \begin{subfigure}[b]{0.4\textwidth}
         \centering
         \includegraphics[width=\textwidth]{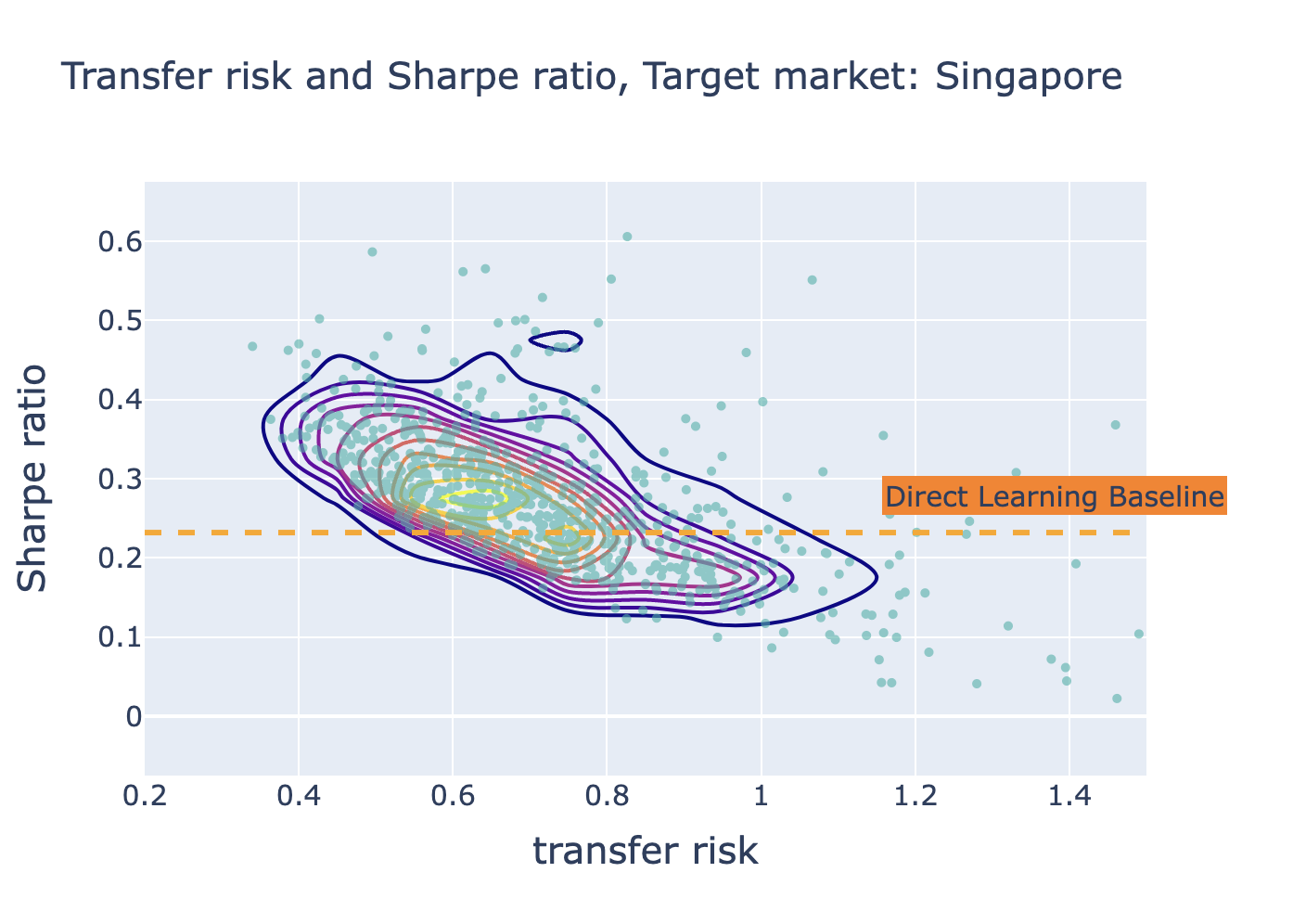}
         \caption{US$\to$Singapore}
         \label{fig:US-Singapore}
     \end{subfigure}
     \caption{Sharpe ratio and transfer risk when transferring from the US market to other markets.}
     \label{fig:cross_continent}
\end{figure}

For each target market, the results across one thousand random experiments (randomness in selections of source assets) are plotted in Figure \ref{fig:cross_continent}. 

\begin{table}[!ht]
  \centering
    \begin{tabular}{c|cccc}
    \toprule
    \toprule
    \textbf{Market} & \textbf{UK} & \textbf{Brazil} & \textbf{Germany} & \textbf{Singapore} \\
    \midrule
    \textbf{Correlation} & -0.66 & -0.67 & -0.64 & -0.62 \\
    \bottomrule
    \bottomrule
    \end{tabular}
  \caption{Correlation between risk and Sharpe ratio for transfer from the US market to other markets.}
  \label{tab:corr_market}
\end{table}

Across those four markets, a consistent pattern is observed: the transfer risk is significantly correlated with the Sharpe ratio of the transferred portfolio (with correlation around -0.60), as shown in Table \ref{tab:corr_market}. This observation supports the idea of using transfer risk as a measurement for the transferability of a task.

Meanwhile, the performance of transfer learning is compared with that of direct learning in Figure \ref{fig:cross_continent}. For each target market, the dashed green line indicates the direct learning performance. The blue dots above the green line represent transfer learning tasks that outperform the direct learning. Note that for all four target markets, there are a significant amount of transfer learning tasks outshining the direct learning, especially for those tasks achieving low transfer risk.

As shown in Figure \ref{fig:cross_continent}, transfer learning is more likely to outperform direct learning in European markets, such as Germany, while it performs worse in the Brazil market. A possible interpretation to this finding is that European markets have tighter connections and higher similarities with the US market, which help to boost the performance of transfer learning.

\subsection{Cross-Sector Transfer.} In this numerical experiment, we focus on transfer learning among ten different sectors in the US equity market: Communication Services, Consumer Discretionary, Energy, Financials, Health Care, Industrials, Information Technology, Materials, Real Estate, and Utilities. We conduct two separate experiments, one with S\&P500 stocks and the other with non-S\&P500 stocks, to compare the differences in transfer learning performance and transfer risk. The analysis reveals that transfer risks in Health Care and Information Technology sectors display large negative correlations with transfer learning outcomes, effectively characterizing transferability in these sectors. In contrast, correlations are not significant for Utilities and Real Estate, potentially due to factors not fully captured by transfer risk. Additionally, for some sectors such as Energy, the correlations become more significant within non-S\&P500 stocks than S\&P500 stocks.

More specifically, given a source sector and a target sector, we first randomly sample ten stocks ($d=10$) from each sector, as the source asset class and target asset class. Then, three data sets will be constructed accordingly:

\begin{enumerate}
    \item \textbf{Source training data}: it consists of the daily returns of ten stocks from a given source sector, from February 2000 to February 2020.

    \item \textbf{Target training data}: it consists of the daily returns of ten stocks from a given target sector, from February 2015 to February 2020.

    \item \textbf{Target testing data}: it consists of the daily returns of ten stocks from a given target sector, from February 2020 to September 2021.
\end{enumerate}

The transfer learning scheme applied in the experiments is same as before: the portfolio is first pre-trained on the source training data by solving \eqref{eqn:portfolio_opt_source}, then fine-tuned on the target training data by solving \eqref{eqn:portfolio_opt_target}. Finally, the performance of the portfolio is evaluated through its Sharpe ratio on the target testing data. The regularization parameter $\lambda$ in \eqref{eqn:portfolio_opt_target} is set to be $0.2$. Meanwhile, the computation of transfer risk follows the approach described in Section \ref{subsec:port_opt_setup}, using the source training data and target testing data. 

\begin{figure}
    \centering
    \includegraphics[width=0.7\textwidth]{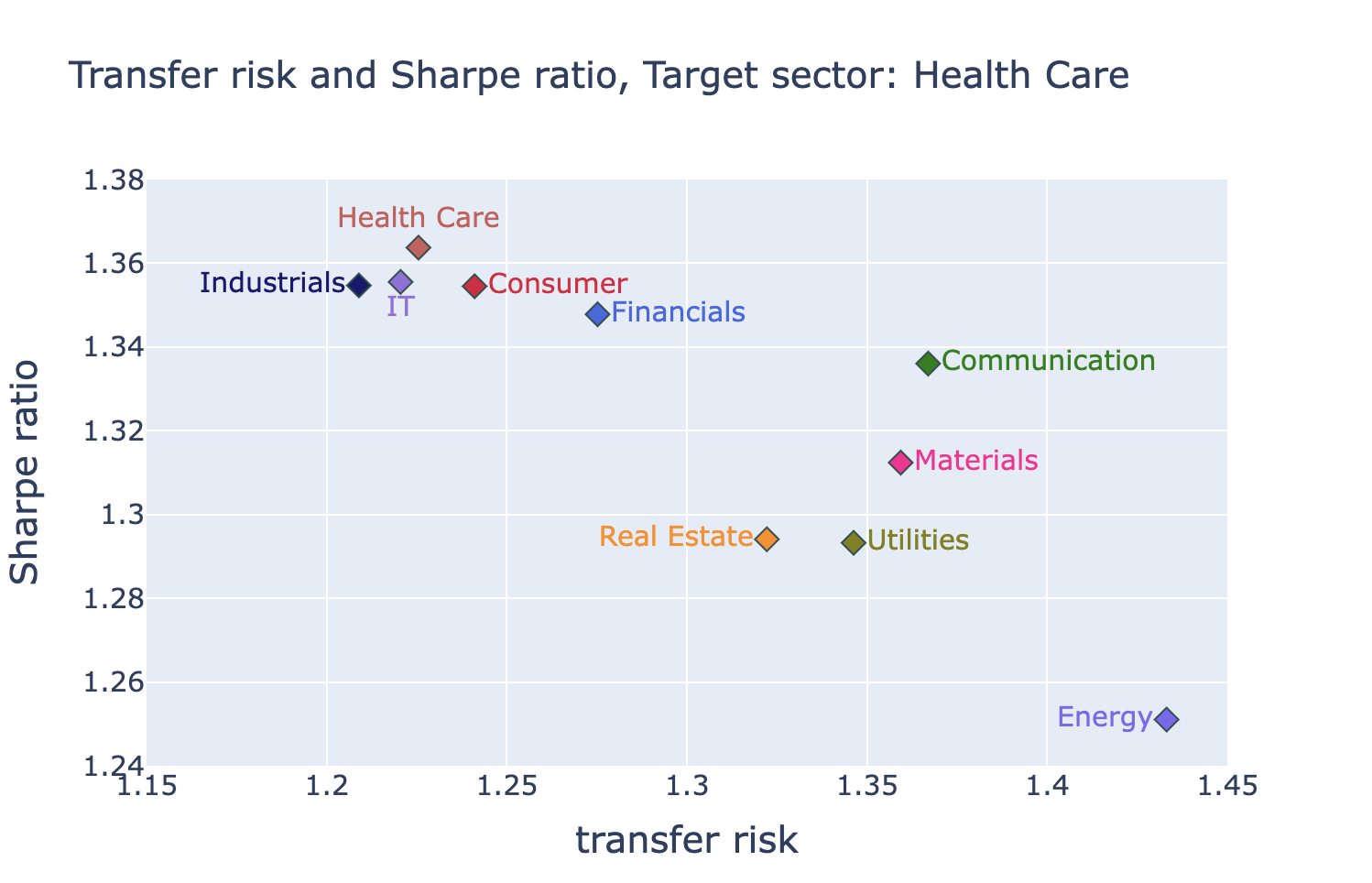}
    \caption{Transfer risk and Sharpe ratio, transferring to Health Care sector in S\&P500 stocks.}
    \label{fig:sector_sp500_health}
\end{figure}

\begin{figure}[ht]
     \centering
     \begin{subfigure}[b]{0.49\textwidth}
         \centering
         \includegraphics[width=\textwidth]{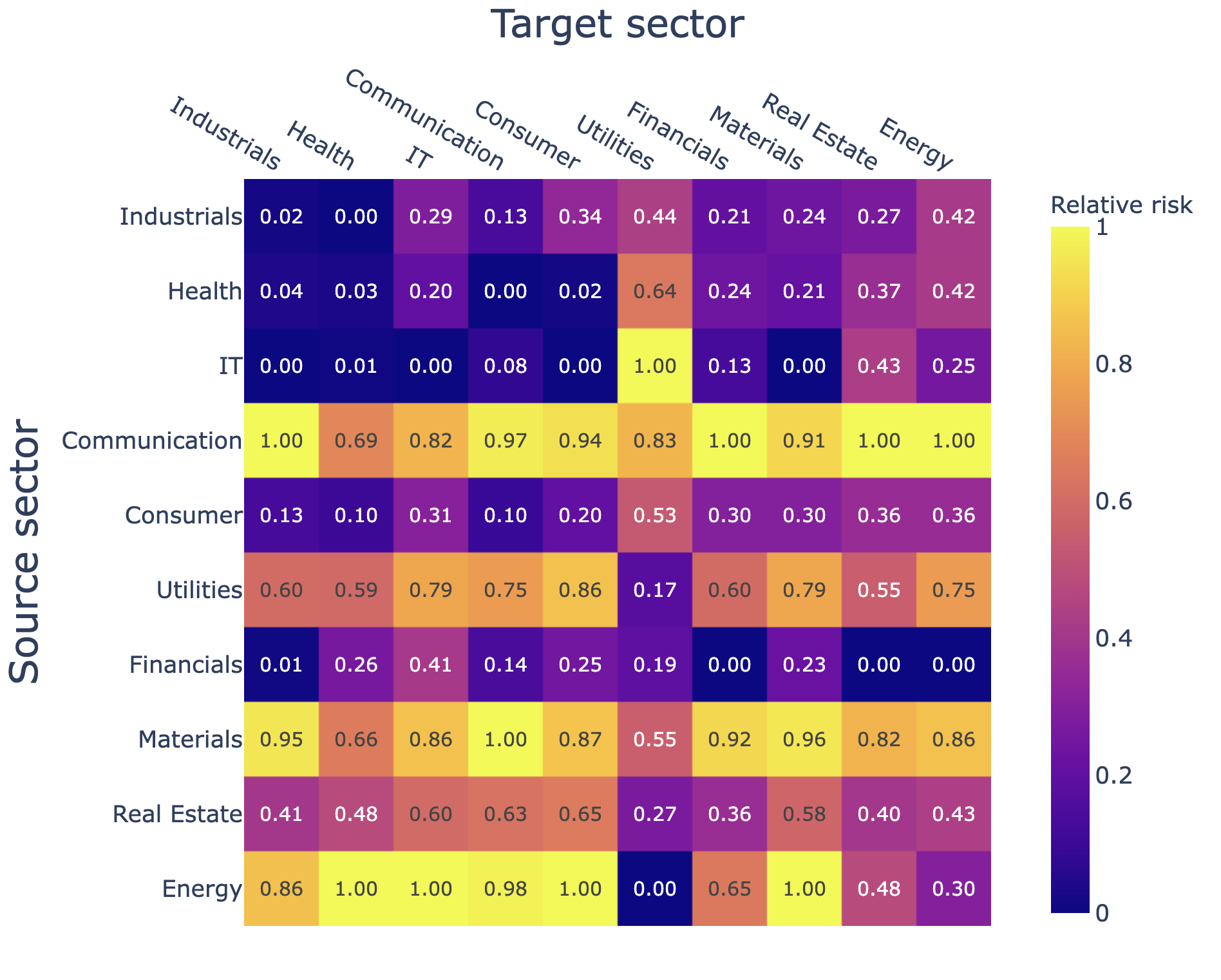}
         \caption{Risk for cross-sector transfer.}
         \label{fig:score_sp500}
     \end{subfigure}
     \hfill
     \begin{subfigure}[b]{0.49\textwidth}
         \centering
         \includegraphics[width=\textwidth]{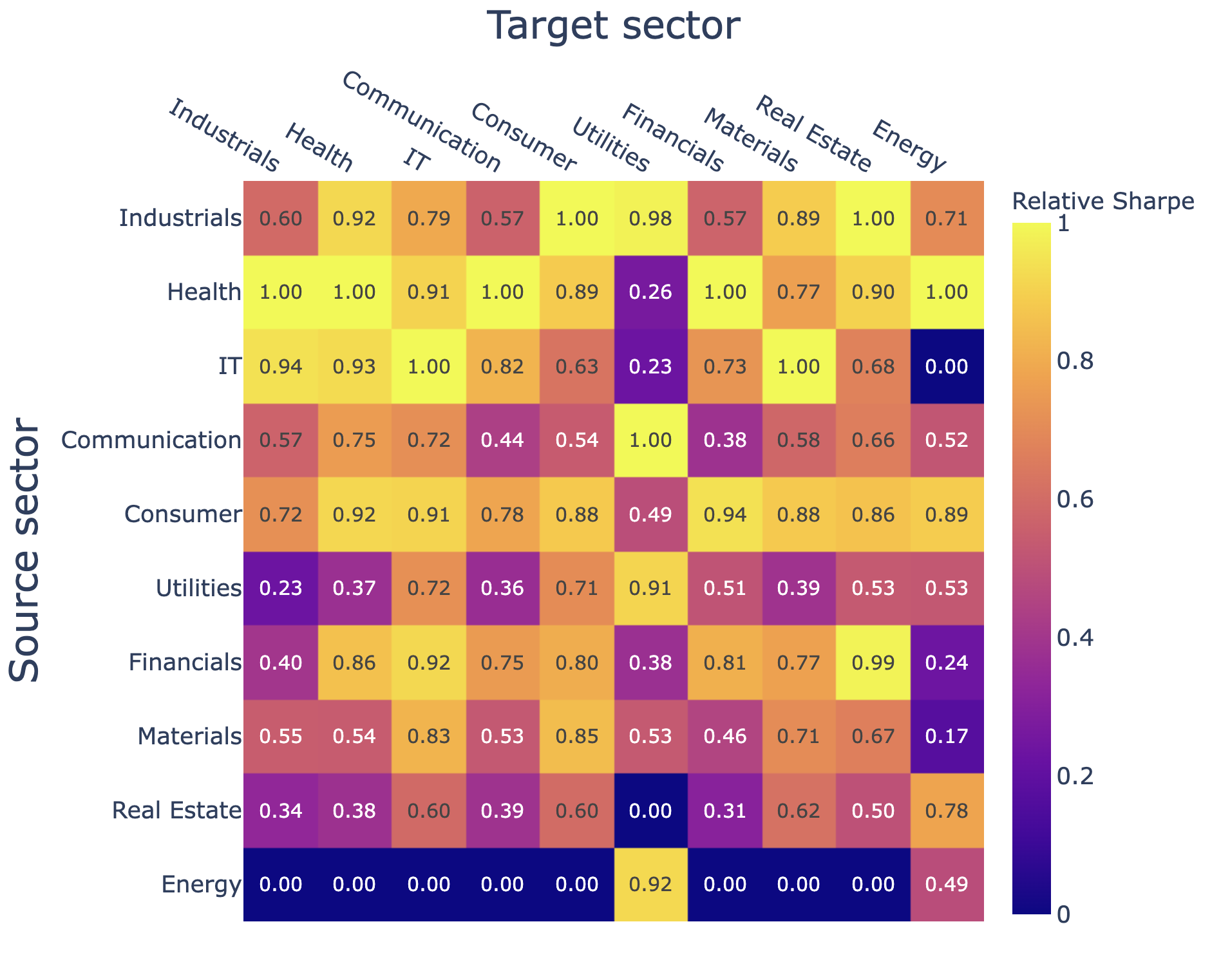}
         \caption{Sharpe ratio for cross-sector transfer.}
         \label{fig:sharpe_sp500}
     \end{subfigure}
     \caption{Relative risk and Sharpe ratio for transfer across sectors with S\&P500 stocks.}
     \label{fig:cross_sector_sp500}
\end{figure}

\begin{figure}[ht]
     \centering
     \begin{subfigure}[b]{0.49\textwidth}
         \centering
         \includegraphics[width=\textwidth]{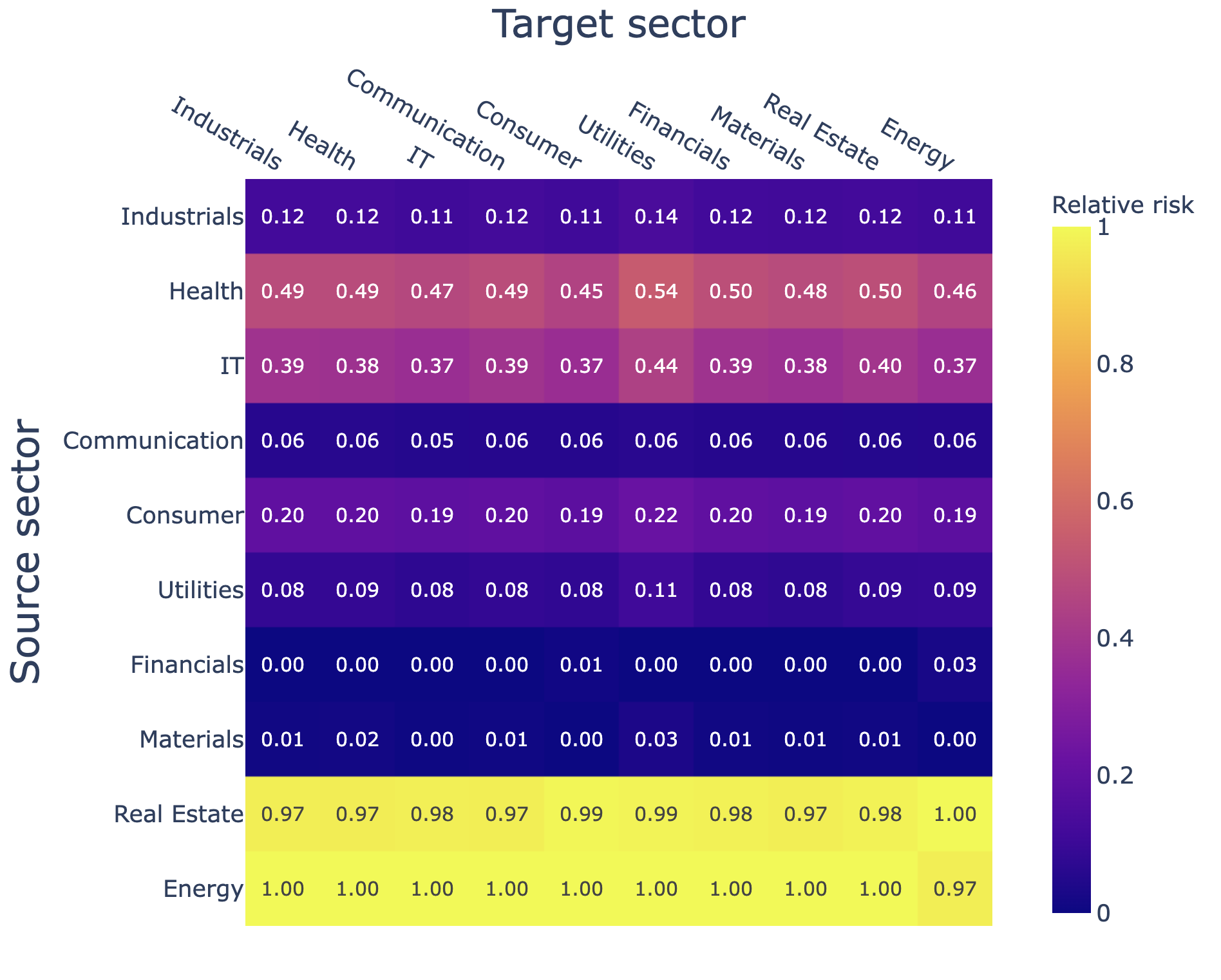}
         \caption{Risk for cross-sector transfer.}
         \label{fig:score_nonsp500}
     \end{subfigure}
     \hfill
     \begin{subfigure}[b]{0.49\textwidth}
         \centering
         \includegraphics[width=\textwidth]{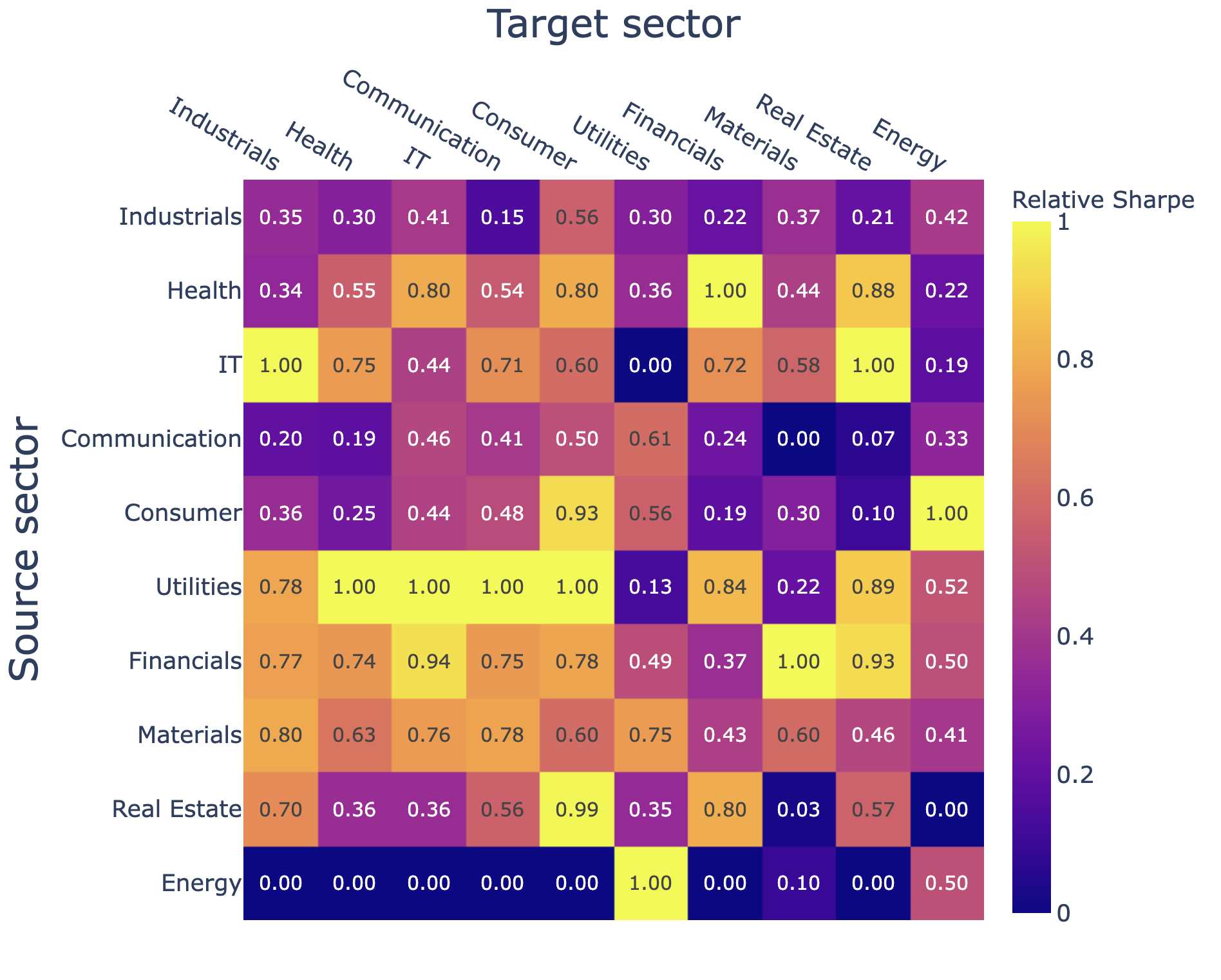}
         \caption{Sharpe ratio for cross-sector transfer.}
         \label{fig:sharpe_nonsp500}
     \end{subfigure}
     \caption{Relative risk and Sharpe ratio for transfer across sectors with non-S\&P500 stocks.}
     \label{fig:cross_sector_nonsp500}
\end{figure}

For each source-target sector pair (in total $10\times10=100$ pairs), five hundred random experiments with different stock selections are conducted, and we record the average Sharpe ratio and average transfer risk of those random experiments. The results are presented in Figure \ref{fig:sector_sp500_health}, Figure \ref{fig:cross_sector_sp500}, Figure \ref{fig:cross_sector_nonsp500} and Table \ref{tab:cross_sector_corr}. 

Figure \ref{fig:sector_sp500_health} shows the relation between (average) Sharpe ratios and (average) scores when transferring portfolios from various source sectors to the target Health Care sector. In general, the negative correlation between Sharpe ratios and scores is observed: when the target sector is fixed (Health Care in this example), transferring a portfolio from a source sector with lower transfer risk is more likely to achieve a higher Sharpe ratio. In particular, Information Technology, Health Care, and Consumer Discretionary are desirable source sectors when the target sector is Health Care, while Energy is not a suitable choice.

In Figure \ref{fig:cross_sector_sp500} and Figure \ref{fig:cross_sector_nonsp500}, we plot the heat maps of transfer risks and Sharpe ratios when transferring across all sectors. Here the transfer risks and Sharpe ratios are rescaled linearly so that for each target sector, the values will range from zero to one, where zero (resp. one) is for the source sector with the lowest (resp. highest) transfer risk or Sharpe ratio. Meanwhile, for experiments in Figure \ref{fig:cross_sector_sp500}, the source and target stocks are all sampled from S\&P500 components, while for Figure \ref{fig:cross_sector_nonsp500}, the source stocks are sampled from stocks not listed in the S\&P500 Index (and the target stocks are still sampled from S\&P500 components).

Table \ref{tab:cross_sector_corr} records the correlation between Sharpe ratios and transfer risks when transferring portfolios from various source sectors to each target sector. Lower correlation implies that the target sector may benefit more from transfer learning and the transferability is better captured by transfer risk. Similar as the setting in Figure \ref{fig:cross_sector_sp500} and Figure \ref{fig:cross_sector_sp500}, Table \ref{tab:sp500_corr} is from the experiments using S\&P500 source stocks, while Table \ref{tab:nonsp500_corr} is from the experiments using non-S\&P500 source stocks.

\begin{table}[ht]
    \begin{subtable}[!ht]{0.49\textwidth}
        \centering
        \begin{tabular}{cc}
            \toprule
            \toprule
            \textbf{Target Sector} & \textbf{Correlation} \\
            \midrule
            \textbf{Health Care} & -0.88 \\
            \textbf{Communication} & -0.82 \\
            \textbf{Materials} & -0.78 \\
            \textbf{IT} & -0.71 \\
            \textbf{Financials} & -0.60 \\
            \textbf{Consumer} & -0.56 \\
            \textbf{Industrials} & -0.54 \\
            \textbf{Real Estate} & -0.37 \\
            \textbf{Utilities} & -0.20 \\
            \textbf{Energy} & 0.04 \\
            \bottomrule
            \bottomrule
        \end{tabular}
       \caption{S\&P500 stocks.}
       \label{tab:sp500_corr}
    \end{subtable}
    \hfill
    \begin{subtable}[!ht]{0.49\textwidth}
        \centering
        \begin{tabular}{cc}
            \toprule
            \toprule
            \textbf{Target Sector} & \textbf{Correlation} \\
            \midrule
            \textbf{IT} & -0.67 \\
            \textbf{Materials} & -0.49 \\
            \textbf{Health Care} & -0.47 \\
            \textbf{Communication} & -0.46 \\
            \textbf{Energy} & -0.40 \\
            \textbf{Industrials} & -0.30 \\
            \textbf{Consumer} & -0.28 \\
            \textbf{Real Estate} & -0.14 \\
            \textbf{Financials} & 0.10 \\
            \textbf{Utilities} & 0.17 \\
            \bottomrule
            \bottomrule
        \end{tabular}
        \caption{Non-S\&P500 stocks.}
        \label{tab:nonsp500_corr}
     \end{subtable}
     \caption{Correlation between risk and Sharpe ratio for transfer from other sectors to the target.}
     \label{tab:cross_sector_corr}
\end{table}

Several insightful patterns are observed from above. For sectors such as Health Care and Information Technology, large negative correlations are revealed in Table \ref{tab:cross_sector_corr}, no matter whether the stocks are chosen from S\&P500 or not. This implies that the transfer risk appropriately encodes the statistical property and characterizes the transferability of portfolios in those sectors. On the contrary, for sectors such as Utilities and Real Estate, the correlations shown in Table \ref{tab:cross_sector_corr} are not significant, regardless of whether the stocks are chosen from S\&P500 or not. This may result from the fact that companies in Utilities and Real Estate sectors tend to be affected by underlying spatial factors and also changes in regulatory policies. Those aspects may not be fully captured by the transfer risk. In addition, for Energy sector, the correlation is more significant when considering non-S\&P500 stocks. This may due to the industry concentration of Energy sector in S\&P500 Index: Energy sector in S\&P500 Index is highly concentrated to a few large companies, and the portfolio's performance is largely driven by some company-specific factors which the transfer risk fails to capture. The effect of industry concentration dwindles when non-S\&P500 stocks are considered.

\subsection{Cross-Frequency Transfer.} In the following numerical experiments, we focus on transfer learning between different trading frequencies, ranging from mid-frequency to low-frequency: 1-minute, 5-minute, 10-minute, 30-minute, 65-minute, 130-minute and 1-day. The findings demonstrate that transferring a low-frequency portfolio (1-day) to higher frequencies results in relatively high transfer risks and poor transfer learning performances. This discrepancy arises from the distinct statistical properties of intraday price movements in mid/high-frequency trading compared to cross-day price movements in low-frequency trading, affecting the transfer process. Conversely, within the mid/high-frequency regime (1-minute to 130-minute), the study reveals that 65-minute and 130-minute frequencies serve as better candidates for the source frequency due to more robust mean and covariance estimations. Consequently, these frequencies lead to improved transfer learning performance after fine-tuning.

More specifically, given a source frequency and a target frequency, we first randomly sample ten stocks ($d=10$) from the fifty largest US companies by market capitalization. Then, three data sets will be constructed accordingly:

\begin{enumerate}
    \item \textbf{Source training data}: it consists of the returns of ten stocks, sampled under a given source frequency, from February 2016 to September 2019.

    \item \textbf{Target training data}: it consists of the returns of ten stocks, sampled under a given target frequency, from February 2016 to September 2019.

    \item \textbf{Target testing data}: it consists of the returns of ten stocks, sampled under a given target frequency, from September 2019 to February 2020.
\end{enumerate}

The transfer learning scheme applied in the experiments is the same as before: the portfolio is first pre-trained by solving \eqref{eqn:portfolio_opt_source} with the mean and covariance estimated from the source-frequency training data, then fine-tuned by solving \eqref{eqn:portfolio_opt_target} with the mean and covariance estimated from the target-frequency training data. Meanwhile, following the usual setting in mid/high-frequency trading, we assume that over-night holding is not allowed for trading frequencies ranging from 1-minute to 130-minute. More specifically, when over-night holding not is allowed, the price movement after the market close and before the market open will not be included in the mean and covariance estimation. Finally, the performance of the portfolio is evaluated through its Sharpe ratio on the target-frequency testing data. The regularization parameter $\lambda$ in \eqref{eqn:portfolio_opt_target} is set to be $0.2$. Meanwhile, the computation of transfer risk follows the approach described in Section \ref{subsec:port_opt_setup}, using the source-frequency training data and target-frequency testing data. 

\begin{figure}
    \centering
    \includegraphics[width=0.7\textwidth]{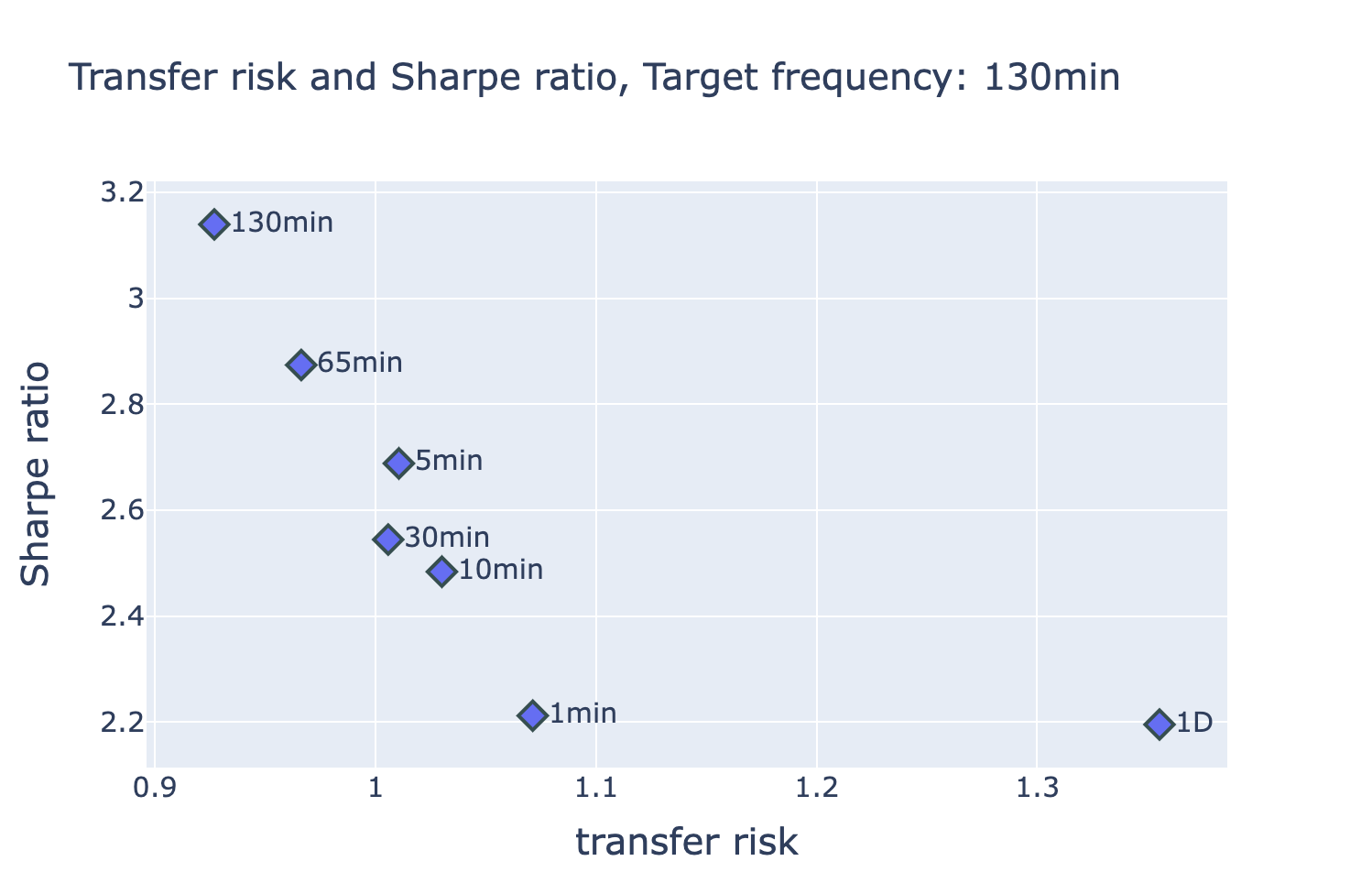}
    \caption{Transfer risk and Sharpe ratio, transferring to 130-minute frequency.}
    \label{fig:freq_intraday_130min}
\end{figure}

\begin{figure}[ht]
     \centering
     \begin{subfigure}[b]{0.49\textwidth}
         \centering
         \includegraphics[width=\textwidth]{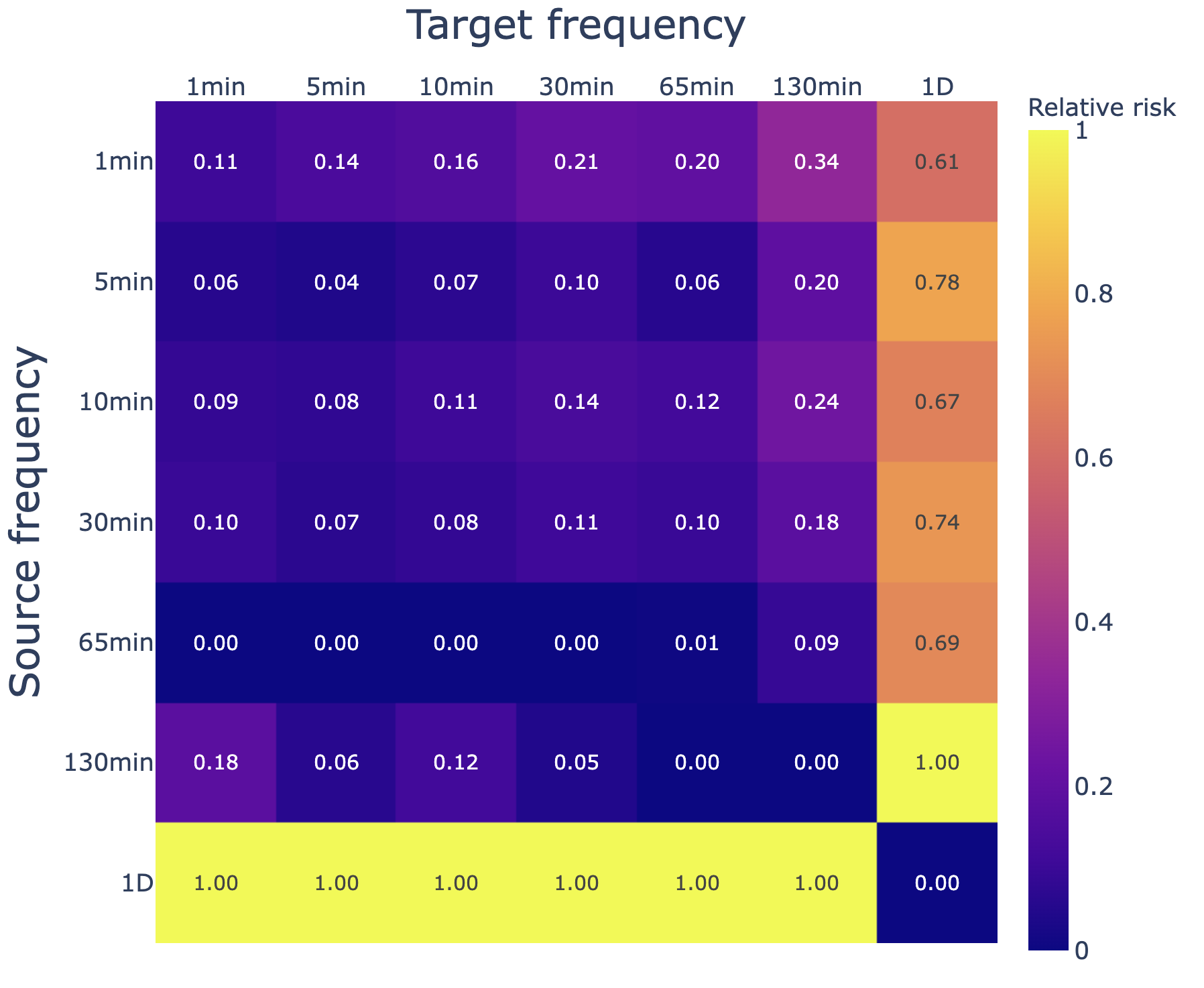}
         \caption{Risk for cross-frequency transfer.}
         \label{fig:score_intraday}
     \end{subfigure}
     \hfill
     \begin{subfigure}[b]{0.49\textwidth}
         \centering
         \includegraphics[width=\textwidth]{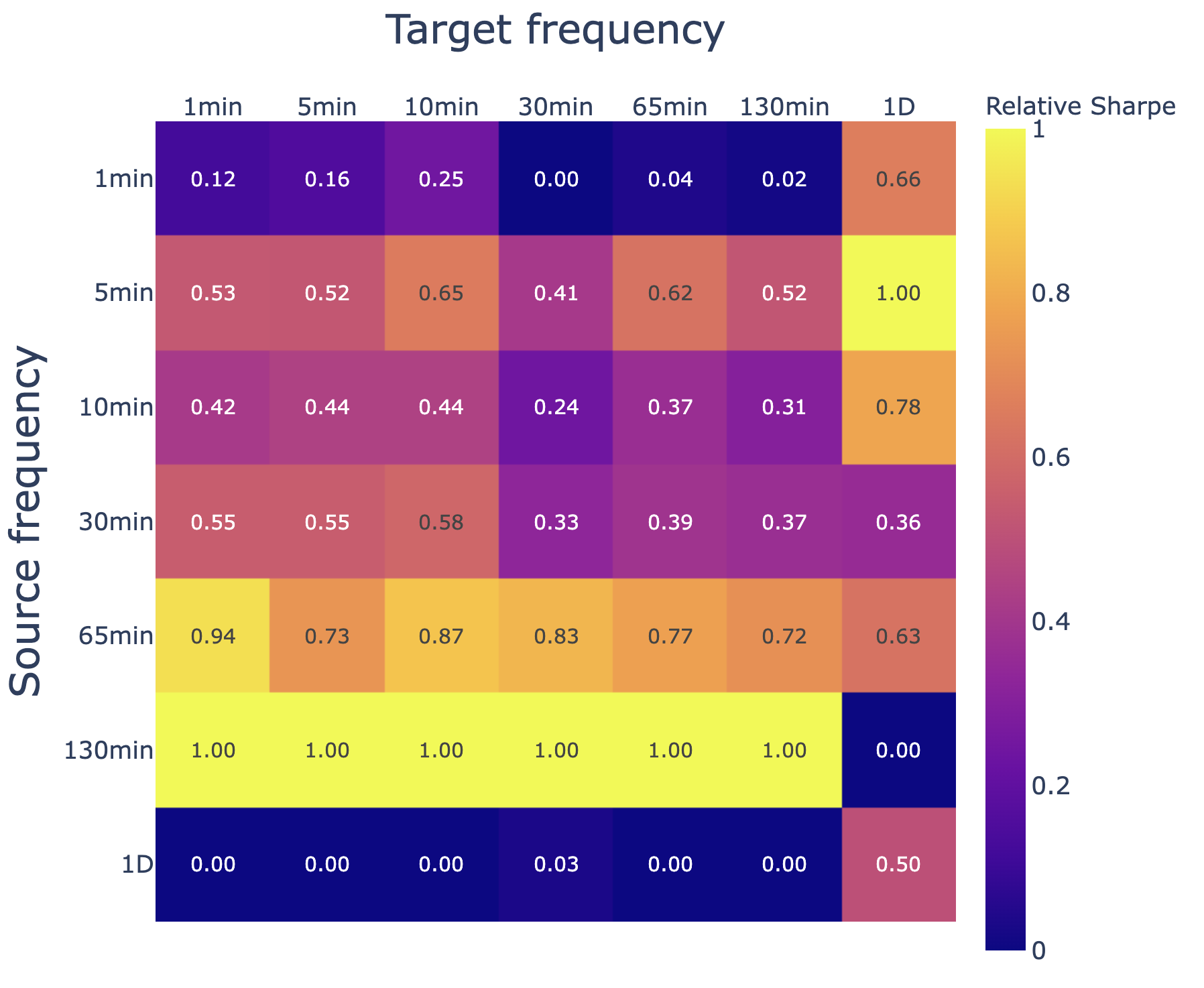}
         \caption{Sharpe ratio for cross-frequency transfer.}
         \label{fig:sharpe_intraday}
     \end{subfigure}
     \caption{Relative risk and Sharpe ratio for transfer across frequencies.}
     \label{fig:cross_freq_intraday}
\end{figure}

For each source-target frequency pair (in total $7\times17=49$ pairs), two hundred random experiments with different stock selections are conducted, and we record the average Sharpe ratio and average transfer risk of those random experiments. The results are presented in Figure \ref{fig:freq_intraday_130min}, Figure \ref{fig:cross_freq_intraday} and Table \ref{tab:cross_freq_corr}. 

For example, Figure \ref{fig:freq_intraday_130min} shows the relation between (average) Sharpe ratios and (average) transfer risks when transferring portfolios from various source frequencies to the target frequency of 130-minute. In general, the negative correlation between Sharpe ratios and scores is observed: when the target frequency is fixed (130-minute in this example), transferring a portfolio from a source frequency with lower transfer risk corresponds to a higher Sharpe ratio. In particular, source frequencies such as 130-minute and 65-minute, which are closer to the 130-minute target frequency, are desirable source tasks, while 1-minute or 1-day is less suitable.

To see more clearly the relation between different frequencies, in Figure \ref{fig:cross_freq_intraday}, we plot the heat maps of transfer risks and Sharpe ratios when transferring across all frequencies. Here the transfer risks and Sharpe ratios are again rescaled linearly, so that for each target frequency, the values will range from zero to one. Meanwhile, Table \ref{tab:cross_freq_corr} records the correlation between Sharpe ratios and transfer risks when transferring portfolios from various source frequencies to each target frequency, following the same setting as Figure \ref{fig:cross_freq_intraday}.

\begin{table}[ht]
    \centering
    \begin{tabular}{cc}
        \toprule
        \toprule
        \textbf{Target Frequency} & \textbf{Correlation} \\
        \midrule
        \textbf{1 min} & -0.59 \\
        \textbf{5 min} & -0.70 \\
        \textbf{10 min} & -0.74 \\
        \textbf{30 min} & -0.59 \\
        \textbf{65 min} & -0.70 \\
        \textbf{130 min} & -0.76 \\
        \textbf{1 Day} & -0.19 \\
        \bottomrule
        \bottomrule
    \end{tabular}
    \caption{Correlation between risk and Sharpe ratio for transfer from other frequency to the target.}
    \label{tab:cross_freq_corr}
\end{table}

From Figure \ref{fig:cross_freq_intraday}, it is observed that the transfer risks from 1-day frequency to other higher frequencies are relatively high, resulting in poor transfer learning performances as well. This demonstrates a natural discrepancy between low-frequency trading and mid/high-frequency trading: mid/high-frequency trading aims to capture intraday stock price movements by not allowing over-night holding, while low-frequency trading intends to capture price movements across trading days. Consequently, the difference between the underlying statistical properties of intraday price movements and cross-day price movements hurts the performance of transferring a low-frequency portfolio to a mid/high-frequency portfolio. 

Meanwhile, for transfer learning inside the mid/high-frequency regime (1-minute to 130-minute), the results in Figure \ref{fig:cross_freq_intraday} reveal that 65-minute and 130-minute are more appropriate candidates for the source frequency, since they lead to much better transfer learning performance, compared to other higher source frequencies. This may due to the fact that under 65-minute and 130-minute frequencies, the mean and covariance estimations are more robust, hence resulting in a robust source portfolio which performs well after fine-tuning.

\section{Significance}\label{sec:conclusion}

In this work, we have demonstrated how to apply transfer learning techniques to solve portfolio optimization problems by leveraging knowledge obtained from well-established portfolios. We have also shown that prior to starting a full-scale transfer learning scheme, transfer risk is an easy-to-compute and viable quantity as a prior estimate of the final learning outcome. By presenting these compelling results, we provide valuable insights into the potential of transfer learning in solving financial problems, with transfer risk as a pivotal component for successful implementation.

\bibliographystyle{apalike}
\bibliography{refs}

\begin{thebibliography}{}

\bibitem[Bao et~al., 2019]{bao2019information}
Bao, Y., Li, Y., Huang, S.-L., Zhang, L., Zheng, L., Zamir, A., and Guibas, L.
  (2019).
\newblock An information-theoretic approach to transferability in task transfer
  learning.
\newblock In {\em 2019 IEEE International Conference on Image Processing},
  pages 2309--2313. IEEE.

\bibitem[Ben-David et~al., 2010]{ben2010theory}
Ben-David, S., Blitzer, J., Crammer, K., Kulesza, A., Pereira, F., and Vaughan,
  J.~W. (2010).
\newblock A theory of learning from different domains.
\newblock {\em Machine learning}, 79(1):151--175.

\bibitem[Blitzer et~al., 2007]{blitzer2007learning}
Blitzer, J., Crammer, K., Kulesza, A., Pereira, F., and Wortman, J. (2007).
\newblock Learning bounds for domain adaptation.
\newblock In {\em Proceedings of the 20th International Conference on Neural
  Information Processing Systems}, volume~20, page 129–136. Curran Associates
  Inc.

\bibitem[Bu et~al., 2020]{bu2020tightening}
Bu, Y., Zou, S., and Veeravalli, V.~V. (2020).
\newblock Tightening mutual information-based bounds on generalization error.
\newblock {\em IEEE Journal on Selected Areas in Information Theory},
  1(1):121--130.

\bibitem[Cao et~al., 2023]{CGG2023}
Cao, H., Gu, H., Guo, X., and Rosenbaum, M. (2023).
\newblock Feasibility and transferability of transfer learning: A mathematical
  framework.
\newblock {\em arXiv preprint arXiv:2301.11542}.

\bibitem[Cen and Wang, 2019]{cen2019crude}
Cen, Z. and Wang, J. (2019).
\newblock Crude oil price prediction model with long short term memory deep
  learning based on prior knowledge data transfer.
\newblock {\em Energy}, 169:160--171.

\bibitem[Cook et~al., 2013]{cook2013transfer}
Cook, D., Feuz, K.~D., and Krishnan, N.~C. (2013).
\newblock Transfer learning for activity recognition: A survey.
\newblock {\em Knowledge and Information Systems}, 36:537--556.

\bibitem[Courty et~al., 2017]{flamary2016optimal}
Courty, N., Flamary, R., Tuia, D., and Rakotomamonjy, A. (2017).
\newblock Optimal transport for domain adaptation.
\newblock {\em IEEE Transactions on Pattern Analysis and Machine Intelligence},
  39(9):1853–1865.

\bibitem[Deng et~al., 2009]{deng2009imagenet}
Deng, J., Dong, W., Socher, R., Li, L.-J., Li, K., and Fei-Fei, L. (2009).
\newblock Image{N}et: A large-scale hierarchical image database.
\newblock In {\em Proceedings of the IEEE Conference on Computer Vision and
  Pattern Recognition}, pages 248--255. IEEE.

\bibitem[Deng et~al., 2013]{deng2013sparse}
Deng, J., Zhang, Z., Marchi, E., and Schuller, B. (2013).
\newblock Sparse autoencoder-based feature transfer learning for speech emotion
  recognition.
\newblock In {\em Proceedings of the 2013 Humaine Association Conference on
  Affective Computing and Intelligent Interaction}, pages 511--516. IEEE.

\bibitem[Devlin et~al., 2019]{devlin-etal-2019-bert}
Devlin, J., Chang, M.-W., Lee, K., and Toutanova, K. (2019).
\newblock {BERT}: Pre-training of deep bidirectional transformers for language
  understanding.
\newblock In {\em Proceedings of the 2019 Conference of the North {A}merican
  Chapter of the Association for Computational Linguistics: Human Language
  Technologies}, volume~1, pages 4171--4186. Association for Computational
  Linguistics.

\bibitem[Ganin and Lempitsky, 2015]{ganin2015unsupervised}
Ganin, Y. and Lempitsky, V. (2015).
\newblock Unsupervised domain adaptation by backpropagation.
\newblock In {\em Proceedings of the 32nd International Conference on Machine
  Learning}, volume~37, pages 1180--1189. PMLR.

\bibitem[Ganin et~al., 2016]{ganin2016domain}
Ganin, Y., Ustinova, E., Ajakan, H., Germain, P., Larochelle, H., Laviolette,
  F., Marchand, M., and Lempitsky, V. (2016).
\newblock Domain-adversarial training of neural networks.
\newblock {\em Journal of Machine Learning Research}, 17(59):1--35.

\bibitem[Harremo{\"e}s and Vajda, 2011]{harremoes2011pairs}
Harremo{\"e}s, P. and Vajda, I. (2011).
\newblock On pairs of $ f $-divergences and their joint range.
\newblock {\em IEEE Transactions on Information Theory}, 57(6):3230--3235.

\bibitem[Hwang and Kuang, 2010]{hwang2010heterogeneous}
Hwang, T. and Kuang, R. (2010).
\newblock A heterogeneous label propagation algorithm for disease gene
  discovery.
\newblock In {\em Proceedings of the 2010 SIAM International Conference on Data
  Mining}, pages 583--594. SIAM.

\bibitem[Jiang and Zhai, 2007]{jiang2007instance}
Jiang, J. and Zhai, C. (2007).
\newblock Instance weighting for domain adaptation in nlp.
\newblock In {\em Proceedings of the 45th Annual Meeting of the Association of
  Computational Linguistics}, pages 264--271.

\bibitem[Kim et~al., 2022]{kim2022transfer}
Kim, H.~E., Cosa-Linan, A., Santhanam, N., Jannesari, M., Maros, M.~E., and
  Ganslandt, T. (2022).
\newblock Transfer learning for medical image classification: A literature
  review.
\newblock {\em BMC Medical Imaging}, 22(1):69.

\bibitem[Leal et~al., 2022]{leal2020learning}
Leal, L., Lauri{\`e}re, M., and Lehalle, C.-A. (2022).
\newblock Learning a functional control for high-frequency finance.
\newblock {\em Quantitative Finance}, 22(11):1973--1987.

\bibitem[Lebichot et~al., 2020]{lebichot2020deep}
Lebichot, B., Le~Borgne, Y.-A., He-Guelton, L., Obl{\'e}, F., and Bontempi, G.
  (2020).
\newblock Deep-learning domain adaptation techniques for credit cards fraud
  detection.
\newblock In {\em Recent Advances in Big Data and Deep Learning: Proceedings of
  the 2019 INNS Big Data and Deep Learning Conference}, pages 78--88. Springer.

\bibitem[Liu et~al., 2019]{liu2019survey}
Liu, R., Shi, Y., Ji, C., and Jia, M. (2019).
\newblock A survey of sentiment analysis based on transfer learning.
\newblock {\em IEEE Access}, 7:85401--85412.

\bibitem[Liu et~al., 2021]{liu2021finbert}
Liu, Z., Huang, D., Huang, K., Li, Z., and Zhao, J. (2021).
\newblock Finbert: A pre-trained financial language representation model for
  financial text mining.
\newblock In {\em Proceedings of the Twenty-Ninth International Joint
  Conference on Artificial Intelligence}, IJCAI'20.

\bibitem[Long et~al., 2015]{long2015learning}
Long, M., Cao, Y., Wang, J., and Jordan, M. (2015).
\newblock Learning transferable features with deep adaptation networks.
\newblock In {\em Proceedings of the 32nd International Conference on Machine
  Learning}, volume~37, pages 97--105. PMLR.

\bibitem[Long et~al., 2014]{long2014transfer}
Long, M., Wang, J., Ding, G., Sun, J., and Yu, P.~S. (2014).
\newblock Transfer joint matching for unsupervised domain adaptation.
\newblock In {\em Proceedings of the IEEE Conference on Computer Vision and
  Pattern Recognition}, pages 1410--1417. IEEE.

\bibitem[OpenAI, 2023]{openai2023gpt4}
OpenAI (2023).
\newblock Gpt-4 technical report.

\bibitem[Ouyang et~al., 2022]{ouyang2022training}
Ouyang, L., Wu, J., Jiang, X., Almeida, D., Wainwright, C., Mishkin, P., Zhang,
  C., Agarwal, S., Slama, K., Ray, A., et~al. (2022).
\newblock Training language models to follow instructions with human feedback.
\newblock {\em Advances in Neural Information Processing Systems},
  35:27730--27744.

\bibitem[Pan and Yang, 2010]{survey1}
Pan, S.~J. and Yang, Q. (2010).
\newblock A survey on transfer learning.
\newblock {\em IEEE Transactions on Knowledge and Data Engineering},
  22(10):1345--1359.

\bibitem[Rosenbaum and Zhang, 2021]{rosenbaum2021deep}
Rosenbaum, M. and Zhang, J. (2021).
\newblock Deep calibration of the quadratic rough heston model.
\newblock {\em arXiv preprint arXiv:2107.01611}.

\bibitem[Ruder et~al., 2019]{ruder2019transfer}
Ruder, S., Peters, M.~E., Swayamdipta, S., and Wolf, T. (2019).
\newblock Transfer learning in natural language processing.
\newblock In {\em Proceedings of the 2019 Conference of the North American
  Chapter of the Association for Computational Linguistics: Tutorials}, pages
  15--18. Association for Computational Linguistics.

\bibitem[Saenko et~al., 2010]{saenko2010adapting}
Saenko, K., Kulis, B., Fritz, M., and Darrell, T. (2010).
\newblock Adapting visual category models to new domains.
\newblock In {\em Proceedings of the 11th European Conference on Computer
  Vision}, pages 213--226. Springer.

\bibitem[Sung et~al., 2022]{sung2022vl}
Sung, Y.-L., Cho, J., and Bansal, M. (2022).
\newblock Vl-adapter: Parameter-efficient transfer learning for
  vision-and-language tasks.
\newblock In {\em Proceedings of the IEEE Conference on Computer Vision and
  Pattern Recognition}, pages 5227--5237. IEEE.

\bibitem[Tan et~al., 2018]{tan2018survey}
Tan, C., Sun, F., Kong, T., Zhang, W., Yang, C., and Liu, C. (2018).
\newblock A survey on deep transfer learning.
\newblock In {\em International Conference on Artificial Neural Networks},
  pages 270--279. Springer.

\bibitem[Tan et~al., 2021]{tan2021otce}
Tan, Y., Li, Y., and Huang, S.-L. (2021).
\newblock O{TCE}: A transferability metric for cross-domain cross-task
  representations.
\newblock In {\em Proceedings of the IEEE Conference on Computer Vision and
  Pattern Recognition}, pages 15779--15788. IEEE.

\bibitem[Tong et~al., 2021]{tong2021mathematical}
Tong, X., Xu, X., Huang, S.-L., and Zheng, L. (2021).
\newblock A mathematical framework for quantifying transferability in
  multi-source transfer learning.
\newblock In {\em Proceedings of the 35th International Conference on Neural
  Information Processing Systems}, volume~34, pages 26103--26116. Curran
  Associates, Inc.

\bibitem[Tzeng et~al., 2017]{tzeng2017adversarial}
Tzeng, E., Hoffman, J., Saenko, K., and Darrell, T. (2017).
\newblock Adversarial discriminative domain adaptation.
\newblock In {\em Proceedings of the IEEE Conference on Computer Vision and
  Pattern Recognition}, pages 7167--7176. IEEE.

\bibitem[Wang et~al., 2022]{wang2022transfer}
Wang, G., Kikuchi, Y., Yi, J., Zou, Q., Zhou, R., and Guo, X. (2022).
\newblock Transfer learning for retinal vascular disease detection: A pilot
  study with diabetic retinopathy and retinopathy of prematurity.
\newblock {\em arXiv preprint arXiv:2201.01250}.

\bibitem[Wang et~al., 2018]{wang2018stratified}
Wang, J., Chen, Y., Hu, L., Peng, X., and Philip, S.~Y. (2018).
\newblock Stratified transfer learning for cross-domain activity recognition.
\newblock In {\em Proceedings of the 2013 {IEEE} International Conference on
  Pervasive Computing and Communications}, pages 1--10. IEEE.

\bibitem[Wang and Deng, 2018]{wang2018deep}
Wang, M. and Deng, W. (2018).
\newblock Deep visual domain adaptation: A survey.
\newblock {\em Neurocomputing}, 312:135--153.

\bibitem[Wu et~al., 2022]{wu2022jointly}
Wu, D., Wang, X., and Wu, S. (2022).
\newblock Jointly modeling transfer learning of industrial chain information
  and deep learning for stock prediction.
\newblock {\em Expert Systems with Applications}, 191:116257.

\bibitem[Xing et~al., 2018]{xing2018natural}
Xing, F.~Z., Cambria, E., and Welsch, R.~E. (2018).
\newblock Natural language based financial forecasting: a survey.
\newblock {\em Artificial Intelligence Review}, 50(1):49--73.

\bibitem[Zeng et~al., 2019]{zeng2019automatic}
Zeng, M., Li, M., Fei, Z., Yu, Y., Pan, Y., and Wang, J. (2019).
\newblock Automatic icd-9 coding via deep transfer learning.
\newblock {\em Neurocomputing}, 324:43--50.

\bibitem[Zhang et~al., 2018]{zhang2018equity}
Zhang, L., Zhang, H., and Hao, S. (2018).
\newblock An equity fund recommendation system by combing transfer learning and
  the utility function of the prospect theory.
\newblock {\em The Journal of Finance and Data Science}, 4(4):223--233.

\bibitem[Zhao et~al., 2019]{zhao2019learning}
Zhao, H., Des~Combes, R.~T., Zhang, K., and Gordon, G. (2019).
\newblock On learning invariant representations for domain adaptation.
\newblock In {\em Proceedings of the 36th International Conference on Machine
  Learning}, volume~97, pages 7523--7532. PMLR.

\bibitem[Zhuang et~al., 2020]{zhuang2020comprehensive}
Zhuang, F., Qi, Z., Duan, K., Xi, D., Zhu, Y., Zhu, H., Xiong, H., and He, Q.
  (2020).
\newblock A comprehensive survey on transfer learning.
\newblock {\em Proceedings of the IEEE}, 109(1):43--76.

\end{thebibliography}
\end{document}